\documentclass[11pt]{article}

\usepackage{cite,times}
\usepackage{color,multicol}
\usepackage [color=yellow]{todonotes}
\usepackage{amsmath,amsthm,amssymb,framed}

\newcommand{\rem}[1]{}

\newcommand{\bu}{\mbox{\boldmath$u$}}

\newcommand{\br}{\mbox{\boldmath$r$}}
\newcommand{\bx}{\mbox{\boldmath$x$}}
\newcommand{\bdf}{\mbox{\boldmath$f$}}

\newcommand{\bel}{\begin{equation}\label}
\newcommand{\ee}{\end{equation}}
\newcommand{\beq}{\begin{eqnarray}\label} 
\newcommand{\eeq}{\end{eqnarray}} 
\newcommand{\bc}{\begin{center}} 
\newcommand{\ec}{\end{center}} 
\newcommand{\ben}{\begin{enumerate}}
\newcommand{\een}{\end{enumerate}}
\newcommand{\bit}{\begin{itemize}}
\newcommand{\eit}{\end{itemize}}

\newcommand\third{\ensuremath{{\scriptstyle\frac{1}{3}}}}
\newcommand\twothirds{\ensuremath{{\scriptstyle\frac{2}{3}}}}

\newcommand\fourfifths{\ensuremath{{\scriptstyle\frac{4}{5}}}}

\newcommand{\Dsd}{d}
\newcommand{\Ren}{\textrm{R}_{\rm e}}
\usepackage{mathtools}

\newtheorem{proposition}{Proposition}[]
\newtheorem{theorem}[proposition]{Theorem}

\newtheorem{lemma}{Lemma}
\theoremstyle{definition}

\numberwithin{equation}{section}
\rem{
\setcounter{section}{0}
\setcounter{subsection}{0}
\setcounter{theorem}{1}
\setcounter{lemma}{1}
\setcounter{corollary}{1}
\setcounter{proposition}{0}
\setcounter{page}{1}

\makeatletter
\@addtoreset{equation}{section}

\makeatother
}

\begin{document}

\begin{center}
\textbf{\large{\color{blue}A correspondence between the multifractal model of turbulence\\
and the Navier-Stokes equations}}
\par\vspace{3mm}\noindent
Berengere Dubrulle,\\
SPEC, CEA, CNRS,\\
Universit\' e Paris-Saclay,\\
F-91191 CEA Paris-Saclay,\\
Gif-sur-Yvette, France.
\par\vspace{3mm}\noindent
and
\par\vspace{5mm}\noindent
J. D. Gibbon,\\
Department of Mathematics,\\
Imperial College London,\\
London SW7 2AZ, UK.
\end{center}


\begin{abstract}
The multifractal model of turbulence (MFM) and the three dimensional Navier-Stokes equations are blended together by applying the probabilistic scaling arguments of the former to a hierarchy of weak solutions of the latter on periodic boundary conditions. This process imposes a lower bound on both the multifractal spectrum $C(h)$, which appears naturally in the Large Deviation formulation of the MFM, and on $h$ the standard scaling parameter. These bounds respectively take the form\,: (i) $C(h) \geq 1-3h$, which is consistent with Kolmogorov's four-fifths law\,; and (ii) $h\geq -\twothirds$. The latter is significant as it prevents solutions from approaching the Navier-Stokes singular set of Caffarelli, Kohn and Nirenberg.
\end{abstract}


\section{\Large Introduction}

In a volume in which a significant number of papers are devoted to turbulent intermittency, it is a moot question whether any correspondence exists between the results derived from fractal theories of turbulence and those derived using the theory of weak solutions of the Navier-Stokes equations. The former uses the methods of fractal physics to determine the statistically steady behaviour of homogeneous, isotropic flows, while the latter uses Sobolev estimates of norms of the velocity field and its derivatives, spatially averaged in a periodic cube, to determine the long-time behaviour of Navier-Stokes solutions. Before studying this correspondence, a brief and mainly descriptive summary of both ideas are given in the rest of this section and the whole of \S\ref{NSEs}.
\par\smallskip
Kolmogorov's 1941 theory, widely known as K41, deals with homogeneous, isotropic flows by using structure function and scaling methods \cite{UF1995}. We begin with the incompressible Euler equations, which are invariant under the scaling transformation 
\bel{scal1}
\bx' = \lambda^{-1}\bx\,,\quad t' = \lambda^{h-1}t\,, \quad \bu' = \lambda^{-h}\bu\,,
\ee
for any value of $h$. For some point $\bx$ in a homogeneous, isotropic flow a sphere of radius $r$ is drawn. Then the scaling in (\ref{scal1}) suggests that the $p$-th order velocity structure function $S_{p}$ should scale like \cite{UF1995}
\bel{Spdef}
S_{p}(r) = \left<\left|\bu(\bx+\br) - \bu(\bx)\right|^{p}\right>_{st.av.} \sim r^{hp}\,.
\ee
K41 showed that the value of $h$ should be fixed at $h=\third$ to ensure that the energy dissipation rate $\varepsilon$ is homogeneous in space and time\footnote{Some of these arguments were anticipated by Onsager in 1949 \cite{Ons,EyinkLN} who posed the question whether energy is conserved in the Euler equations if the data is very rough. Using heuristic arguments he postulated that $h=\third$ is the critical value below which energy conservation does not occur. For modern work on this problem see \cite{deLSz,CET1994,Is2016}.}. Thus we have $S_{p} \sim r^{p/3}$. When $p=3$ an exact relation is recovered where the right hand side of (\ref{Spdef}) is equal to $-\fourfifths\varepsilon r$, with an $O(\nu)$ Navier-Stokes correction. This is called Kolmogorov's four-fifths law \cite{UF1995}.
\par\vspace{1mm}
Parisi and Frisch \cite{PF1985} then introduced an argument that relaxes the enforcement of the value $h=\third$ to allow a continuous spectrum of exponents $h$, based on the scale invariance (\ref{scal1}), provided the energy dissipation rate $\varepsilon$ is constant ``on the average''. Exactly what this average means is determined by the introduction of $P_{r}(h)$, the probability of observing a given scaling exponent $h$ at the scale $r$. In the mulifractal model's original formulation $P_{r}(h)$ was computed by assuming that each value of $h$ belongs to a given fractal set of dimension $D(h)$ \cite{UF1995,BJPV1998,BB2009,PF1985,PV1987,PV1987rev,Boff2008,Frisch2016,BDB2019}. As explained in a set of lecture notes by Eyink \cite{EyinkLN} (see also \cite{BDB2019}) a more precise mathematical definition can be established by using Large Deviation Theory. In this formulation $P_{r}(h)$ is chosen as   
\bel{Prdef}
P_{r}(h)\sim r^{C(h)}\,,
\ee
where $C(h)$ is called the multi-fractal spectrum and has encoded within it all the properties of intermittency in the flow\footnote{In its original interpretation \cite{PF1985,UF1995,EyinkLN}, in a $\Dsd$-dimensional spatial domain, $C(h)$ corresponds to the co-dimension of the fractal set, with a fractal dimension $D(h) = \Dsd\ - C(h)$. The subtle difference is that in the Large Deviation interpretation \cite{EyinkLN,BDB2019}, it is theoretically possible to have $C(h) \geq d$. In \S\ref{hiphip} it is shown that this is indeed the case.}.  The structure functions $S_{p}(r)$, instead of taking their K41-form as in (\ref{Spdef}) with $h=\third$, are now expressed as
\beq{Spzeta}
S_{p}(r) &\sim& r^{\zeta_{p}}\,,\\
\zeta_{p} &=& \inf_{h}\left[hp + C(h)\right]\,.
\eeq
The relation $\zeta_{3}=1$, corresponding to Kolmogorov's four-fifths law, then leads to the constraint $C(h)\ge 1-3h$, equality being reached for a monofractal. In this case the equality $C(h)= 1-3h$, combined with the condition $\zeta_{3}=1$, results in $h = \third$ with probability one. For the more general case where the inequality is strict, a whole family of exponents $h$ can be found with non-zero probability. The value $\zeta_{3}=1$ is achieved for an exponent $h$ shifted from $\third$, which is the hallmark of intermittency. Altogether, this picture has been named the ``multifractal'' approach to intermittency in turbulence.
\par\vspace{1mm}
By blending the probabilistic scaling ideas of the MFM into the weak solution formulation of the Navier-Stokes equations on periodic boundary conditions, including the effect of derivatives to arbitrarily high order, it is shown in \S\ref{hiphip} that the lower bound 
\bel{Cbd1}
C(h)\geq 1-3h\,,
\ee
still holds. Thus the consistency with the four-fifths law holds for arbitrarily high derivatives. This is the first of the two main results of the paper. The second is the existence of a lower bound on $h$ 
\bel{Cbd2}
h \geq (1-d)/3\,,
\ee
which becomes $h\geq -\twothirds$ when $d=3$. This is discussed in \S\ref{CKN} and is of crucial importance\,: there it is suggested that bounding $h$ away from its Leray value of $h=-1$ prevents solutions from approaching the Navier-Stokes singular set of Caffarelli, Kohn and Nirenberg as $r \to 0$ \cite{CKN1982}.

\section{\Large Weak solutions and length scales of the Navier-Stokes equations}\label{NSEs}

\subsection{\large Weak solutions of the three-dimensional Navier-Stokes equations}\label{weak}

Consider a divergence-free velocity field $\bu(\bx,\,t)$ evolving according to the Navier-Stokes equations \cite{Leray1934,FGT1981,DG1995,FMRT2001,DF2002,RRS2016}
\bel{NS1}
\left(\partial_{t}+\bu\cdot\nabla\right)\bu + \nabla p =  \nu\Delta\bu + \bdf(\bx)\,,
\ee
on a three-dimensional periodic domain $V = [0,\,L]^{3}_{per}$, with $\nu$ as the viscosity and $\bdf$ is an $L^{2}$-bounded forcing. We define a doubly-labelled set of norms in dimensionless form
\bel{NS2a}
F_{n,m}(t) = \nu^{-1} L^{1/\alpha_{n,m}} \|\nabla^{n}\bu\|_{2m}\,,
\ee
for $1 \leq n < \infty$ and $1 \leq m \leq \infty$, where
\bel{NS2b}
 \|\nabla^{n}\bu\|_{2m} = \left(\int_{V}|\nabla^{n}\bu|^{2m}dV\right)^{1/2m}\,.
\ee
The $\alpha_{n,m}$ in the exponents of $L^{1/\alpha_{n,m}}$ in (\ref{NS2a}) are defined by 
\bel{alphadef}
\alpha_{n,m} = \frac{2m}{2m(n+1)-3}\,.
\ee
This factor ensures that the $F_{n,m}$ are dimensionless. In addition to the invariant Euler scalings on $(\bx,\,t,\,\bu)$ in (\ref{scal1}), a rescaling of $\nu$ is required for Navier-Stokes invariance\footnote{The choice of $h=-1$ leaves the viscosity invariant and leads to the Leray scaling $\bu(\bx,\,t) \to \lambda^{-1}\bu\left(\bx/\lambda,\,t/\lambda^{2}\right)$ \cite{Leray1934}.} 
\bel{rescal2}
\bx' = \lambda^{-1}\bx\,;\quad t' = \lambda^{h-1}t\,; \quad \bu' = \lambda^{-h}\bu\,; \quad \nu' = \lambda^{-(h+1)}\nu\,.
\ee
It can then be directly verified that the $F_{n,m}(t)$ defined in (\ref{NS2a}) are invariant under the scale invariance property (\ref{rescal2}) for every finite value of the dimensionless parameter $\lambda \neq 0$ and of $h$. This invariance at every length and time scale in the flow makes the set of  $F_{n,m}$ invaluable as a tool for investigating a cascade of energy through the system. The higher derivatives, labelled by $n$, are sensitive to ever finer length scales in the flow while higher values of $m$ pick out the larger spikes, with the $m=\infty$ case representing the maximum norm. 
\par\vspace{1mm}
Before continuing, it is necessary to make a remark about the distinction between the angled brackets $\left<\cdot\right>_{st.av.}$ that denote a statistical average in K41 theory and the angled brackets $\left<\cdot\right>_{T}$ that denote a time average up to time $T>0$ in the time-evolving Navier-Stokes equations. The latter pair of brackets are defined by 
\bel{brackdef}
\left<\cdot\right>_{T} = T^{-1}\int_{0}^{T}\cdot\,dt\,.
\ee
This is used initially in the definition of a space-time averaged velocity $U$ that appears in the Reynolds number $Re$. Their definitions are 
\bel{Redef}
L^{3}U^{2} = \left<\|\bu\|_{2}^{2}\right>_{T}\qquad\mbox{and}\qquad Re = UL\,\nu^{-1}\,.
\ee
As defined in (\ref{Redef}), the Reynolds number $Re$ is the response of the system to the forcing $\bdf(\bx)$. How this is related to both the Grashof number $Gr$ through $\|\bdf\|_{2}$, and to $Re$ itself is explained in \cite{DF2002,JDG2018}. Using both (\ref{brackdef}) and (\ref{Redef}) the first result to note in this paper is that there exists a bounded, weighted, double hierarchy of their time averages proved in \cite{JDG2018}\,:
\begin{theorem}\label{thm1}
For $n \geq  1$ and $1 \leq m \leq \infty$, on periodic boundary conditions, weak solutions of the three-dimensional Navier-Stokes equations obey
\bel{NS3}
\left<F_{n,m}^{\,\alpha_{n,m}}\right>_{T} \leq c_{n,m}Re^{3} + O\left(T^{-1}\right)\,,
\ee
where the $c_{n,m}$ are a set of constants.
\end{theorem}
\par\vspace{-1mm}\noindent
Note that for $n=m=1$ we have the correctly bounded energy dissipation rate for the standard three-dimensional Navier-Stokes equations, from which one can easily obtain a $Re^{3/4}$ bound on the inverse Kolmogorov length $L\lambda_{k}^{-1}$. As noted in \cite{JDG2018}, Theorem \ref{thm1} in its fullest form encapsulates all the known Leray-type weak solution results in Navier-Stokes analysis which are distributional in nature but not unique \cite{Leray1934,FGT1981,DG1995,FMRT2001,DF2002,RRS2016}.  In \cite{JDG2018} it was also shown that to prove full regularity (existence \textit{and} uniqueness of solutions) we would need to prove that
\bel{NS4}
\left<F_{n,m}^{2\alpha_{n,m}}\right>_{T} < \infty\,.
\ee
While this remains an open problem, there is no evidence that any bounds with the factor of 2 in the exponent exist. 


\subsection{\large A hierarchy of Navier-Stokes length scales}

To extract a definition of a set of length scales from the three-dimensional Navier-Stokes equations consider the semi-norm
\bel{ls1a}
H_{n,m} = \int_{\mathcal{V}} |\nabla^{n}\bu|^{2m}\,dV\,.
\ee
Using dimensional analysis we define a set of $t$-dependent length-scales $\{\lambda_{n,m}(t)\}$ such that
\bel{ls1b}
\lambda_{n,m}^{-2m(n+1) + 3}\nu^{2m} = \left(\frac{L}{\lambda_{n,m}}\right)^{-3}H_{n,m}\,.
\ee
The re-scaled inverse volume $L^{-3}$ on the right hand side is inserted to be sure that (\ref{ls1b}) gives the correct definition of the Kolmogorov length when $m=n=1$. Solving (\ref{ls1b}) for $\lambda_{n,m}$ gives 
\bel{ls2}
\left(L\lambda_{n,m}^{-1}\right)^{n+1} = F_{n,m}\,.
\ee
where $\alpha_{n,m}$ is defined in (\ref{alphadef}). This enables us to use Theorem \ref{thm1} to obtain \cite{JDG2018}\,: 
\begin{lemma}\label{lem1}
When $n \geq  1$ and $1 \leq m \leq \infty$ the $\lambda_{n,m}^{-1}(t)$ satisfy the time averages
\bel{lslem1}
\left<\left(L\lambda_{n,m}^{-1}\right)^{(n+1)\alpha_{n,m}}\right>_{T} \leq c_{n,m}Re^{3} + O\left(T^{-1}\right)\,.
\ee
\end{lemma}
\par\vspace{0mm}\noindent
Because $(n+1)\alpha_{n,m} >1$, a Holder inequality applied to the left hand side of (\ref{lslem1}) then gives
\bel{lslem2}
\left<L\lambda_{n,m}^{-1}\right>_{T} \leq c_{n,m}Re^{\frac{3}{(n+1)\alpha_{n,m}}} + O\left(T^{-1}\right)\,.
\ee
The exponent on upper bound on the right hand side of (\ref{lslem2}) has a finite limit\,: 
\bel{ls3}
\lim_{n,m\to \infty}\frac{3}{(n+1)\alpha_{n,m}}\to 3\,,
\ee
thus suggesting that both Richardson and Kolmogorov were correct in their assumption that viscosity finally halts the cascade process\,: see the discussion in \cite{JDG2020}. 

\subsection{\large A Navier-Stokes result in integer $\Dsd$ dimensions}\label{NSDdim}

In \cite{JDG2020} it has been shown how the results of Theorem \ref{thm1} can be generalized to a $\Dsd$-dimensional 
domain\footnote{In \cite{JDG2020} the letter $D$ was used as the dimension of the spatial domain. Here we use $\Dsd$ to avoid confusion with the fractal dimension $D(h)$ in multifractal theory.} for integer values $\Dsd=1,\,2$ and $3$. This involves a generalization of the results of Theorem \ref{thm1} where $F_{n,m,\Dsd}$ is now defined as
\bel{Dim1}
F_{n,m,\Dsd} = \nu^{-1} L^{1/\alpha_{n,m,\Dsd}} \|\nabla^{n}\bu\|_{2m}\,,
\ee
together with a more general definition of $\alpha_{n,m}$ which is now called $\alpha_{n,m,\Dsd}$
\bel{alphaDdef}
\alpha_{n,m,\Dsd} = \frac{2m}{2m(n+1)-\Dsd}\,.
\ee
The $F_{n,m,\Dsd}$ possess the same invariance properties as $F_{n,m}$ in (\ref{NS2a}) \cite{JDG2020}.
\begin{theorem}\label{thmD}
For $\Dsd=2,\,3$, and for $n \geq 1$ and $1 \leq m \leq \infty$
\bel{Dim2}
\left<F_{n,m,\Dsd}^{(4-\Dsd)\alpha_{n,m,\Dsd}}\right>_{T} \leq c_{n,m,\Dsd}\,Re^{3}\,.
\ee
For $\Dsd=1$ the same result holds for Burgers' equation.
\end{theorem}
\par\vspace{-1mm}\noindent
It should be stressed that there are no currently available methods that enable such a result to be proved for non-integer values of $\Dsd$. This issue will be discussed at greater length in \S\ref{con}.


\section{\Large Lower bounds on $h$ and $C(h)$}\label{hiphip}

A correspondence between multifractal theory and the Navier-Stokes equations is more appropriate for the stage when $T$ in the time averages $\left<\cdot\right>_{T}$ in (\ref{lslem2}) and (\ref{Dim2}) is large enough such that a Navier-Stokes turbulent flow has reached the fully developed stage. Multifractal theory enables us to obtain the scaling of $H_{n,m}$, defined in (\ref{ls1a}), as a function of $\nu$ (in the limit $\nu \to 0$) via an $h$-dependent dissipation length scale $\eta_{h}$ defined as that scale at which the local Reynold number $(\delta u) \ell/\nu=1$ \cite{PV1987,PV1987rev}. As explained in  \cite{BDB2019}, this separates the scales into two domains\,; a self-similar domain where $\delta u \sim \ell^h$ for $\ell>\eta_h$ and a laminar domain where $\delta u$ is regular for $\ell<\eta_h$. Thus a balance occurs when 
\bel{etadef1}
\eta_h\sim \nu^{\frac{1}{1+h}}\,.
\ee
Moreover, in the laminar domain, $\delta u$ can be Taylor-expanded, resulting in $\delta u\approx\ell |\nabla\bu|$. Matching with the self-similar domain at $\eta_h$ then provides the scalings
\bel{etadef2}
|\nabla \bu| \sim \eta_h^{h-1}\,,\qquad\mbox{and}\qquad |\nabla^{n}\bu|^{2m}\sim \eta_h^{2m(h-n)}\,.
\ee
Then, forming the correspondence
\bel{classicMFR}
L^{-3}\int_{\mathcal{V}} |\nabla^{n}\bu|^{2m}\,dV  \quad\longleftrightarrow \quad \int_{h} \eta_h^{2m(h-n) }P_{\eta_h}(h)dh\,,
\ee
and using the scalings $\eta_h\sim \nu^{1/(1+h)}$ and $H_{n,m}\sim L^3 \nu^{\chi_{n,m}}$, in the limit $\nu\to 0$, we obtain
\bel{chidemn}
\chi_{n,m}=\min_{h}\left(\frac{2m(h+1)+C(h)-2m(n+1)}{1+h}\right)\,.
\ee
Using the definition of $F_{n,m}$ in (\ref{NS2a}) and comparing this with eq. (\ref{ls2}), we therefore obtain
\bel{newcorr}
\lambda_{n,m}^{-(n+1)}\sim \nu^{\frac{\chi_{n,m}}{2m}-1}\,.
\ee
Inserting this estimate into (\ref{lslem2}) of Lemma \ref{lem1}, and comparing powers of $\nu$, in the limit $\nu\to 0$, leads to the condition
\bel{condition}
\frac{\chi_{n,m}}{2m} - 1\ge  -\frac{3}{\alpha_{n,m}}\,. 
\ee
In dimension $\Dsd$, the value of $\chi_{n,m}$ does not change and we have the more general condition
\bel{conditionD}
\frac{\chi_{n,m}}{2m} - 1 \ge  - \frac{3}{(4-\Dsd)\alpha_{n,m,\Dsd}}=\frac{3}{4-\Dsd}\left[\frac{\Dsd}{2m}-(n+1)\right]\,.
\ee
Developing eqs. (\ref{chidemn}) and (\ref{conditionD}) leads to 
\bel{oufeq}
C(h) \ge 2m(n+1)\left(1-\frac{3(1+h)}{4-\Dsd}\right)+\frac{3\Dsd(1+h)}{4-\Dsd},\quad\forall (n,m)\ge 1\,.
\ee
In the limit $(n,m)\to \infty$ the right hand side of equation (\ref{oufeq}) goes to infinity, unless $h \geq (1-d)/3$, which means that the only scaling exponents that have a nonzero probability are those greater than 
\bel{hmin1}
h_{min}= (1-\Dsd)/3\,.
\ee
When $\Dsd = 3$ we have the lower bound 
\bel{hmin2}
h \geq - 2/3\,,
\ee
the consequences of which will be discussed in \S\ref{CKN}.
\par\smallskip
Returning to (\ref{oufeq}), for any $h\ge h_{min}$, the sharpest bound on $C(h)$, uniform in $n,\,m$ comes from the values $m=n=1$, leading to
\bel{sharpest}
C(h)\ge 1-3h\,,\qquad\mbox{with}\qquad C(h_{min})\ge \Dsd\,.
\ee
The result $C(h_{min})\ge \Dsd$ looks unusual but has a very low probability of occurrence. It is indeed one of the features allowed by Large Deviation Theory \cite{EyinkLN}.
\par\vspace*{0mm}\noindent
\begin{figure}[!h]
\includegraphics[scale=0.33]{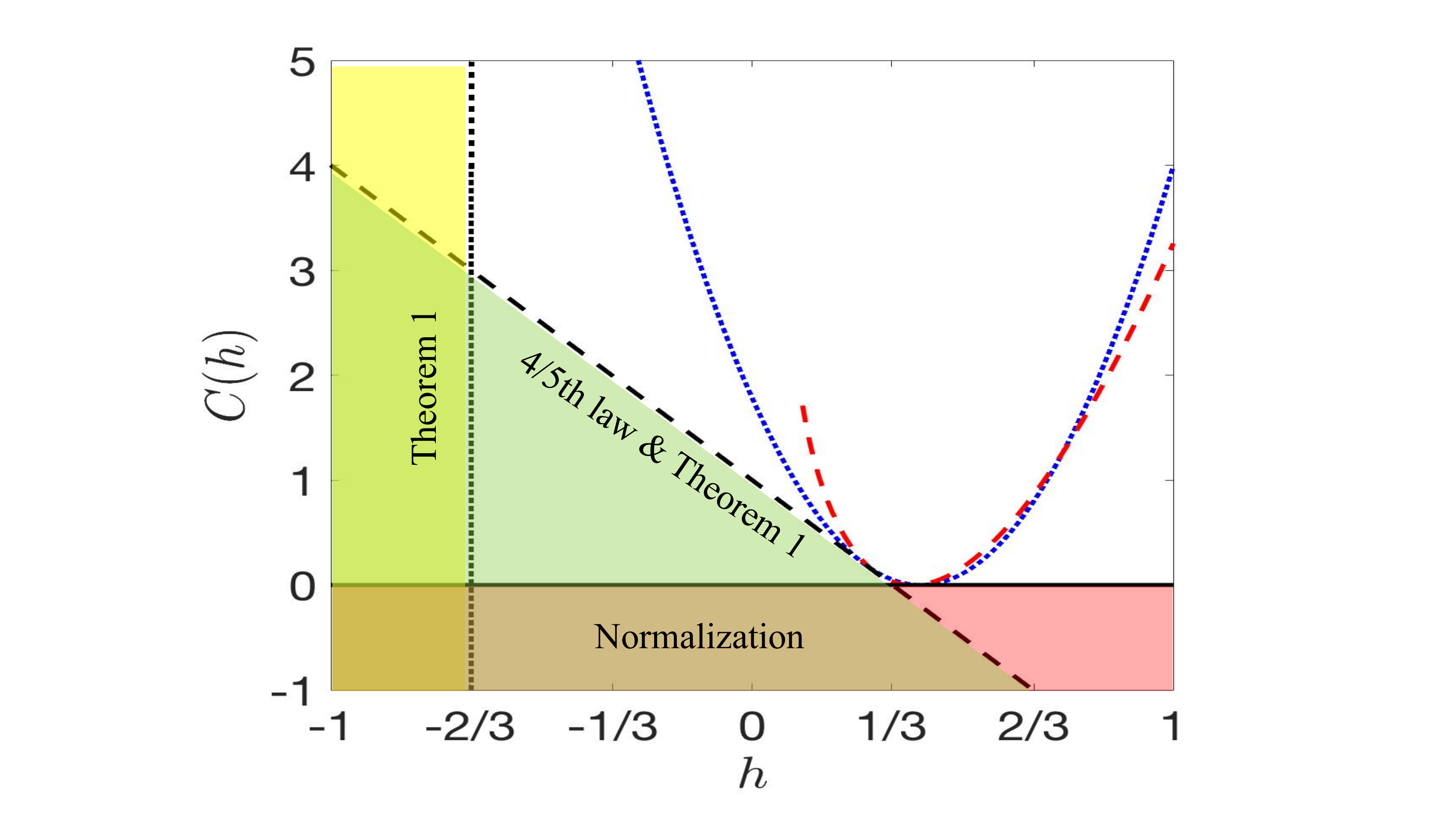}
\caption{\scriptsize The figure is compiled from the various constraints, including $C(h)\geq 1-3h$, and displays the admissibility range of the multifractal spectrum when $\Dsd= 3$. The yellow zone left of the vertical dotted line is excluded as a result of Theorem 1. The red zone below the horizontal continuous line is excluded as a result of normalization of $C(h)$ \cite{BDB2019}. The green zone below the black dashed-dotted line is excluded as a result of both Theorem 1 and the four-fifths law. A few classical models that fit turbulence measurements for $\Dsd=3$ are also shown\,: blue dotted line\,: log-normal model with $b=0.045$ \cite{BDB2019}; red dashed line\,: log-Poisson model with $\beta=2/3$ \cite{SL94}.}\label{fig:conditionsall}
\end{figure}
\par\vspace*{-5mm}
The first inequality in (\ref{sharpest}) is the same as that derived from the four-fifths law \cite{BDB2019}. The second inequality $C(h_{min})\ge \Dsd$ also provides a bound on the probability of observing the smallest exponent which has a very low probability of occurrence \cite{CKN1982}. Note that that for $\Dsd=3$, the condition for $n=m=1$ reflects the fact that the energy dissipation is bounded, as stressed in \cite{JDG2020}. Thus, we can say that while the hierarchy of bounds derived in Theorem 1 gives a lower bound\footnote{Paladin and Vulpiani \cite{PV1987,PV1987rev} introduced the idea of an $h$-dependent dissipation scale $\eta_{h}$ such that $L\eta_{h}^{-1} \sim \Ren^{1/(1+h)}$. Comparing this with the estimate in (\ref{ls2}) requires the assumption that their Reynolds number $\Ren$ is equivalent to $Re$ of (\ref{Redef}).  Assuming this, we have $3(1+h) = (4-d)(n+1)\alpha_{n,m,d}$. Given that $m,\,n$ vary across the ranges $1 \leq m \leq \infty$ and $n\geq 1$ we end up with $(1-\Dsd)/3 \leq h \leq \third$, which is consistent with (\ref{hmin1}).} on $h$, which is expressed in (\ref{hmin1}), the lower bound $C(h)$ in (\ref{sharpest}) is no better than that derived from the four-fifths law. We see also that the exponent corresponding to Leray scaling, namely $h_{min}=-1$, is only achieved at the singular dimension $\Dsd=4$. All the constraints on the multifractal spectrum are assembled in Fig. \ref{fig:conditionsall}. Thanks to this, we can also place it in context the classical multifractal models, such as the log-normal model \cite{K62}
\bel{logN}
C(h)=(h-1/3 -3b/2)^2/2b\,,
\ee
and the log-Poisson model \cite{Boff2008,SL94} 
\bel{logP}
C(h)=-\frac{2\beta-3h-1}{\ln(\beta)}\left[1-\ln\left(\frac{2\beta-3h-1}{2\ln\beta}\right)\right]\,,
\ee
both of which are plotted in Fig. \ref{fig:conditionsall}. 


\section{\Large Singularities or no singularities? The CKN singular set}\label{CKN}

In their influential paper, Caffarelli, Kohn and Nirenberg \cite{CKN1982} developed the work of Scheffer \cite{VS1,VS2,VS3} by showing that for suitable weak solutions of the three-dimensional Navier-Stokes equations, the singular set in space-time has zero one-dimensional Hausdorff measure.  We shall refer to this as the CKN singular set. The scaling invariance (\ref{scal1}) plays a significant role in their proof which uses the technical innovation of covering the CKN set with space-time parabolic cylinders instead of Euclidean balls. Of particular interest is their result\footnote{See Corollary 1 to Proposition 1 in \S1 of \cite{CKN1982}.} which shows that in the limit $r\to 0$, as solutions approach the CKN singular set, the velocity field $\bu$ must obey
\bel{CKN1}
|\bu| > c\,r^{-1},\quad \mbox{as}\quad r\to 0\,,
\ee
where $r^{2}= \left(x-x_{0}\right)^{2}+\nu\left(t-t_{0}\right)$ is the distance from a suitably chosen point $\left(x_{0},\,t_{0}\right)$ on the axis of a space-time parabolic cylinder. The $r^{-1}$ lower bound in (\ref{CKN1}) can be interpreted as a minimal rate of approach to the the CKN singular set for for which the corresponding value of $h$ is $h=-1$. This suggests that the lower bound $h=-\twothirds$ given in (\ref{hmin2}) prevents solutions from approaching this set. An alternative and more general way of expressing this result in $\Dsd$-dimensions is to say that for any $\Dsd < 4$, which implies that $h_{min} = \third (1-d) > -1$, the Leray scaling exponent $h=-1$ has no probability of occurrence. It should be stressed that this is \textit{not} a rigorous proof of Navier-Stokes regularity for $\Dsd=3$\,: the lower bound $h_{min}$ in (\ref{hmin1}) has its origin in the application of the MFM and the dissipation length scale $\eta_{h}$ to the Navier-Stokes equations. This is a collective result derived from blending the two models together. Nevertheless, to a physicist it is highly suggestive that Navier-Stokes singularities are prohibited on periodic boundary conditions. There are parallels with the case of the three dimensional Euler equations, where the formation of a singularity seems to be plausible for flows with boundaries \cite{Hou}, but are still elusive for the case with periodic boundary conditions.

\section{\Large Conclusion}\label{con}

The power of the multifractal method lies in the fact that a spread of values of the parameter $h$ are employed. Whereas the discussion so far has been restricted to the lower bounds on $h$ and $C(h)$, now we wish to discuss the physical manifestations of these bounds. In its original formulation there exists a fractal dimension $D(h)$ for each value of $h$ \cite{UF1995,EyinkLN}. Is there a parallel with the Navier-Stokes equations operating over an equivalent range of dimensions? Before addressing this question, at this stage it is worth considering the implications of the many numerical simulations performed on the three-dimensional Navier-Stokes equations  \cite{DCB1991,TK1993,JWSR1993,VM1994,MKO1994,DYS2008,IGK2009,EM2010,HIK2013,HIWK2014,EIGSH2017,EIH2020}. The contours of the dissipation or strain fields are typically displayed in a cube and all have similar features\,; at intermediate times flattening occurs producing pancake-like structures, followed by the roll-up of these quasi-two-dimensional objects into quasi-one-dimensional, stretched filaments. Depending on both the initial conditions and the Reynolds number, this process may undergo repetition at a local level. Nevertheless, the trend towards filamentation is universal as the turbulence becomes more fully developed. The dimension of the set(s) on which dissipation accumulates appears to drop from $3$ down to $1$ and even below that as the filaments themselves begin to break down.
\par\vspace{1mm}
To put this in its historical context, the idea has lingered for many years that Navier-Stokes solutions might accumulate on an attractor embedded in the full domain-space whose dimension is less than 3. A simple finite dimensional example would be the Lorenz equations whose attractor, embedded in 3-dimensional phase space, has a box-counting dimension of about 2.05. In terms of the issues addressed in this paper the equivalent question to ask is whether Theorem \ref{thmD} could be true for non-integer values of $\Dsd < 3$? However, given the state of current methods in rigorous Navier-Stokes analysis, there is no question of being able to prove Theorem \ref{thmD} when $\Dsd$ is not an integer. Although this idea cannot be entirely ruled out, it is our belief that this is the wrong question to ask. Instead we prefer to look at the result in Theorem \ref{thmD} in the following way. Inequality (\ref{Dim2}) shows how the exponent of $F_{n,m,\Dsd}$ scales with the domain dimension $\Dsd$. When $n=m=1$, the surprising but crucial factor of $4-\Dsd$ cancels to make $(4-\Dsd)\alpha_{1,1,\Dsd}=2$ for every value of $\Dsd$. It also furnishes us with the correct bound on the averaged energy dissipation rate. When $\Dsd=2$ we reach 
\bel{Dim3}
\left[(4-\Dsd)\alpha_{n,m,\Dsd}\right]_{\Dsd=2} = 2\alpha_{n,m,2}\,.
\ee
The factor of $2$ in the upper bound puts it into the category of (\ref{NS4}) thereby confirming that the case $\Dsd=2$ is critical for regularity. Writing out in full the exponent of $F_{n,m,\Dsd}$, one finds that  
\bel{Dim4}
(4-\Dsd)\alpha_{n,m,\Dsd} = \frac{2m(4-\Dsd)}{2m(n+1) - \Dsd}\,,
\ee
which \textit{increases} as $\Dsd\searrow 0$. What we do know is that an increasing exponent of $F_{n,m,\Dsd}$ implies more, not less, regularity, which lies in the direction of \textit{increasing dissipation}. In numerical simulations the apparent drop in the dimension of the sets on which vorticity or dissipation accumulates suggests that a flow may adjust itself to find the smoothest, most dissipative set on which to operate not the most singular, which is behaviour consistent with the avoidance of the singular set demonstrated in \S\ref{CKN}.
\par\smallskip\noindent
\textbf{Acknowledgements\,:} We thank A. Cheskidov, C. Doering, R. Shvydkoy, J. T. Stuart and D. Vincenzi for useful comments.  B.D. acknowledges support from the ANR, project EXPLOIT, grant agreement no. ANR-16-CE06-0006-01. This research was supported in part by the International Centre for Theoretical Sciences (ICTS),  Bengaluru, for the online program {\em Turbulence\,: Problems at the Interface of Mathematics and Physics}, (code: ICTS/TPIMP2020/12).



\begin{thebibliography}{9}\itemsep -1mm
{\small
\bibitem{UF1995} Frisch U. 1995 \textit{Turbulence\,: The Legacy of A. N. Kolmogorov}. Cambridge, UK\,: Cambridge University Press.


\bibitem{Ons} Eyink GL. and Sreenivasan KR. 2006 Onsager and the theory of hydrodynamic turbulence. \textit{Rev. Mod. Phys.} \textbf{78}, 87--135. 

\bibitem{EyinkLN} Eyink GL. 2008 \textit{Turbulence Theory -- unpublished lecture notes chapters 3c and 3d.} The Johns Hopkins University, Baltimore. See http://www.ams.jhu.edu/$\sim$eyink/Turbulence/notes/  

\bibitem{deLSz}  De Lellis C. and Szekelyhidi L. 2009 The Euler equations as a differential inclusion. \textit{Ann. Math.} \textbf{170}, 1417--1436.

\bibitem{CET1994} Constantin P., E W. and Titi ES. 1994 Onsager's conjecture on the energy conservation for solutions of Euler's equation.  \textit{Comm. Math. Phys.} \textbf{165}, 207--209. 

\bibitem{Is2016} Isett P. 2018 A proof of Onsager's Conjecture. \textit{Ann. Math.} \textbf{188:3}, 1--93.

\bibitem{PF1985}  Parisi G. and Frisch U. 1985 In\,: Ghil M., Benzi R. and Parisi G. (eds.) \textit{Turbulence and Predictability in Geophysical Fluid Dynamics}. Proc. Int. School of Physics ``E. Fermi'', pp. 84--87. Amsterdam, NL. North-Holland.

\bibitem{BJPV1998} Bohr T., Jensen MH., Paladin G. and Vulpiani A. 1998 \textit{Dynamical Systems Approach to Turbulence}. Cambridge, UK\,: Cambridge University Press. 

\bibitem{BB2009} Benzi R. and Biferale L. 2009 Fully Developed Turbulence and the Multifractal Conjecture. \textit{J. Stat. Phys.} \textbf{135}, 977--990. 


\bibitem{PV1987} Paladin G. and Vulpiani A. 1987 Degrees of freedom in turbulence. \textit{Phys. Rev. A} \textbf{35}, R1971. 

\bibitem{PV1987rev} Paladin G. and Vulpiani A. 1987 Anomalous scaling laws in fractal objects. \textit{Physics Reports} \textbf{156}, No. 4, 147--225.

\bibitem{Boff2008} Boffetta G., Mazzino A. and Vulpiani A. 2008 Twenty-five years of multifractals in fully developed turbulence\,: a tribute to Giovanni Paladin. \textit{J. Phys. A} \textbf{41},  363001.

\bibitem{Frisch2016} Frisch U. 2016 The collective birth of multifractals. \textit{J. Phys. A: Math. Theoret.} \textbf{49}, 451002.

\bibitem{BDB2019} Dubrulle B. 2019 Beyond Kolmogorov cascades. \textit{J. Fluid Mech.} \textbf{867}, P1--63.


\bibitem{CKN1982} Caffarelli L., Kohn R. and Nirenberg L. 1982 Partial regularity of suitable weak solutions of the Navier-Stokes equations. \textit{Comm. Pure Appl. Math.} \textbf{35}, 771--831. 

\bibitem{Leray1934} Leray J. 1934 Sur le mouvement d'un liquide visqueux emplissant l'\'espace. \textit{Acta Math.} \textbf{63}, 193--248. 

\bibitem{FGT1981} Foias C., Guillop\'e C. and Temam R. 1981 New a priori estimates for Navier-Stokes equations in dimension 3. \textit{Com. Part. Diff. Equns.} \textbf{6}(3), 329--359.
    
\bibitem{DG1995} Doering CR. and Gibbon JD. 1995 \textit{Applied Analysis of the Navier-Stokes Equations}. Cambridge, UK: Cambridge University Press. 

\bibitem{FMRT2001} Foias C., Manley O., Rosa R. and Temam R. 2001 \textit{Navier-Stokes Equations and Turbulence}. Cambridge, UK: Cambridge University Press. 

\bibitem{DF2002} Doering CR. and Foias C. 2002 Energy dissipation in body-forced turbulence. \textit{J. Fluid Mech.} \textbf{467}, 289--306. 

\bibitem{RRS2016} Robinson JC. Rodrigo JL. and Sadowski W. 2016 \textit{The Three-dimensional Navier-Stokes Equations\,: Classical Theory}\,;  Cambridge Studies in Advanced Mathematics. Cambridge, UK: Cambridge University Press.  

\bibitem{JDG2018} Gibbon JD. 2019 Weak and Strong Solutions of the 3D Navier-Stokes Equations and Their Relation to a Chessboard of Convergent Inverse Length Scales. \textit{J. Nonlin. Sci.} \textbf{29}(1), 215--228. 

\bibitem{JDG2020} Gibbon JD. 2020 Turbulent cascades and thin sets in $3D$ NS-turbulence. \textit{EPL} \textbf{131}, 64001. 

\bibitem{K62} Kolmogorov AN. 1962 A refinement of previous hypotheses concerning the local structure of turbulence in a viscous incompressible fluid at high Reynolds number. \textit{J. Fluid Mech.} \textbf{13}, 82--85.
  
\bibitem{SL94} She ZS. and Leveque E. 1994 Universal scaling laws in fully developed turbulence. \textit{Phys. Rev. Lett.} \textbf{72}, 3, 336--339.

\bibitem{VS1} Scheffer V. 1976 Partial regularity of solutions to the Navier-Stokes equations. \textit{Pacific J. Math.} \textbf{66}, 535--552.

\bibitem{VS2} Scheffer, V. 1977 Hausdorff measure and the Navier-Stokes equations. \textit{Comm. Math. Phys.} \textbf{55}, 97--112.

\bibitem{VS3} Scheffer V. 1978 The Navier-Stokes equations in space dimension four. \textit{Comm. Math. Phys.} \textbf{61}, 41--68.

\bibitem{Hou} Luo G. and  Hou TY. 2014  Potentially singular solutions of the 3D axisymmetric Euler equations. \textit{Proc. Natl. Acad. Sci. USA}  \textbf{111,}  12968--12973.


\bibitem{DCB1991} Douady S., Couder Y. and Brachet ME. 1991 Direct observation of the intermittency of intense vorticity filaments in turbulence.  \textit{Phys. Rev. Lett.} \textbf{67}, 983--986.

\bibitem{TK1993} Tanaka M. and Kida S. 1993 Characterization of vortex tubes and sheets. \textit{Phys. Fluids} \textbf{5}, 2079--2082.

\bibitem{JWSR1993} Jimenez J. Wray A., Saffman PG. and Rogallo RS. 1993 The structure of intense vorticity in isotropic turbulence. 
\textit{J. Fluid Mech.} \textbf{255}, 65--90.

\bibitem{VM1994} Vincent A. and Meneguzzi M. 1994 The dynamics of vorticity tubes of homogeneous turbulence. \textit{J. Fluid Mech.} \textbf{225}, 245--254.

\bibitem{MKO1994} Moffatt HK., Kida S. and Ohkitani K. 1994 Stretched vortices --- the sinews of turbulence\,; large-Reynolds-number asymptotics. \textit{J. Fluid Mech.} \textbf{259}, 241--264.

\bibitem{DYS2008} Donzis D., Yeung PK. and Sreenivasan KR. 2008 Dissipation and enstrophy in isotropic turbulence\,: scaling and resolution effects in direct numerical simulations. \textit{Phys. Fluids} \textbf{20}, 045108.

\bibitem{IGK2009} Ishihara T., Gotoh T. and Kaneda Y. 2009 Study of high-Reynolds number isotropic turbulence by direct numerical simulation. \textit{Annu. Rev. Fluid Mech.} \textbf{41}, 165--180.

\bibitem{EM2010} Elsinga GE. and Marusic I. 2010 Universal aspects of small-scale motions in turbulence. \textit{J. Fluid Mech.} \textbf{662}, 514--539.

\bibitem{HIK2013} Hunt JCR., Ishihara T., Worth NA. and Kaneda Y. 2013 Thin Shear Layers in High Reynolds Number Turbulence -- DNS Results.  \textit{Flow, Turbulence and Combustion} \textbf{91}, 895--929. 

\bibitem{HIWK2014} Hunt JCR., Ishihara T., Worth NA. and Kaneda Y. 2014 Thin shear layer structures in high Reynolds number turbulence. \textit{Flow, Turbulence and Combustion} \textbf{92}, 607--649.

\bibitem{EIGSH2017} Elsinga GE., Ishihara T., Goudar MV., da Silva CB. and Hunt JCR. 2017 The scaling of straining motions in homogeneous isotropic turbulence. \textit{J. Fluid Mech.} \textbf{829}, 61--64.

\bibitem{EIH2020} Elsinga GE., Ishihara T. and Hunt JCR. 2020 Extreme dissipation and intermittency in turbulence at very high Reynolds numbers.  \textit{Proc. R. Soc. A} \textbf{476}, 20200591.


}
\end{thebibliography}
\end{document}